\documentclass[aps,pr,twocolumn,superscriptaddress,showpacs,preprintnumbers,amsmath,amssymb]{revtex4}
\usepackage{graphicx} 
\usepackage{dcolumn}  

\usepackage{epsfig}
\usepackage{refmerge}

\def \vec #1{\mbox{{\boldmath $#1$}}}

\def \lr #1{\left( #1 \right)}
\def \suplr #1{^{\left( #1 \right)}}

\def \Diff {{\mathit \Delta}}

\def \abs #1{\left| #1 \right|}

\def \F {{\cal F}}

\def \GeV {{\rm GeV}}
\def \MeV {{\rm MeV}}

\def \eV {{\rm eV}}

\begin{document}



%
\title{ 
High statistics study of the $f_0\lr{980}$ resonance
in $\gamma\gamma\to\pi^+\pi^-$ production
}

\begin{abstract}
We report on a high statistics measurement of the cross section
of the process $\gamma\gamma\to\pi^+\pi^-$
in the $\pi^+\pi^-$ invariant mass range $0.8~\GeV/c^2 < W < 1.5~\GeV/c^2$
with 85.9~fb$^{-1}$ of data collected at $\sqrt{s}=10.58$~GeV
and 10.52~GeV with the Belle detector.
A clear signal for the $f_0(980)$ resonance is observed.
From a fit to the mass spectrum, the mass, $\pi^+\pi^-$ and
two-photon decay widths of the resonance are found to be 
$985.6 ~^{+1.2}_{-1.5}\lr{\rm stat}
                  ~^{+1.1}_{-1.6}\lr{\rm syst}~\MeV/c^2,\; 
34.2~^{+13.9}_{-11.8}\lr{\rm stat}
                             ~^{+8.8}_{-2.5}\lr{\rm syst}~\MeV,\; {\rm and}
\;\; 205 ~^{+95}_{-83}\lr{\rm stat}
                                   ~_{-117}^{+147}\lr{\rm syst}$ eV,
respectively.
\end{abstract}


\affiliation{Budker Institute of Nuclear Physics, Novosibirsk}
\affiliation{Chiba University, Chiba}
\affiliation{Chonnam National University, Kwangju}
\affiliation{University of Cincinnati, Cincinnati, Ohio 45221}
\affiliation{Department of Physics, Fu Jen Catholic University, Taipei}
\affiliation{The Graduate University for Advanced Studies, Hayama, Japan}
\affiliation{University of Hawaii, Honolulu, Hawaii 96822}
\affiliation{High Energy Accelerator Research Organization (KEK), Tsukuba}
\affiliation{Hiroshima Institute of Technology, Hiroshima}
\affiliation{University of Illinois at Urbana-Champaign, Urbana, Illinois 61801}
\affiliation{Institute of High Energy Physics, Chinese Academy of Sciences, Beijing}
\affiliation{Institute of High Energy Physics, Vienna}
\affiliation{Institute of High Energy Physics, Protvino}
\affiliation{Institute for Theoretical and Experimental Physics, Moscow}
\affiliation{J. Stefan Institute, Ljubljana}
\affiliation{Kanagawa University, Yokohama}
\affiliation{Korea University, Seoul}
\affiliation{Kyungpook National University, Taegu}
\affiliation{Swiss Federal Institute of Technology of Lausanne, EPFL, Lausanne}
\affiliation{University of Ljubljana, Ljubljana}
\affiliation{University of Maribor, Maribor}
\affiliation{University of Melbourne, Victoria}
\affiliation{Nagoya University, Nagoya}
\affiliation{Nara Women's University, Nara}
\affiliation{National Central University, Chung-li}
\affiliation{National United University, Miao Li}
\affiliation{Department of Physics, National Taiwan University, Taipei}
\affiliation{H. Niewodniczanski Institute of Nuclear Physics, Krakow}
\affiliation{Nippon Dental University, Niigata}
\affiliation{Niigata University, Niigata}
\affiliation{University of Nova Gorica, Nova Gorica}
\affiliation{Osaka City University, Osaka}
\affiliation{Osaka University, Osaka}
\affiliation{Panjab University, Chandigarh}
\affiliation{Peking University, Beijing}
\affiliation{RIKEN BNL Research Center, Upton, New York 11973}
\affiliation{University of Science and Technology of China, Hefei}
\affiliation{Seoul National University, Seoul}
\affiliation{Shinshu University, Nagano}
\affiliation{Sungkyunkwan University, Suwon}
\affiliation{University of Sydney, Sydney NSW}
\affiliation{Tata Institute of Fundamental Research, Bombay}
\affiliation{Toho University, Funabashi}
\affiliation{Tohoku Gakuin University, Tagajo}
\affiliation{Tohoku University, Sendai}
\affiliation{Department of Physics, University of Tokyo, Tokyo}
\affiliation{Tokyo Institute of Technology, Tokyo}
\affiliation{Tokyo Metropolitan University, Tokyo}
\affiliation{Tokyo University of Agriculture and Technology, Tokyo}
\affiliation{Virginia Polytechnic Institute and State University, Blacksburg, Virginia 24061}
\affiliation{Yonsei University, Seoul}
  \author{T.~Mori}\affiliation{Nagoya University, Nagoya} 
  \author{S.~Uehara}\affiliation{High Energy Accelerator Research Organization (KEK), Tsukuba} 
  \author{Y.~Watanabe}\affiliation{Tokyo Institute of Technology, Tokyo} 
  \author{K.~Abe}\affiliation{High Energy Accelerator Research Organization (KEK), Tsukuba} 
  \author{K.~Abe}\affiliation{Tohoku Gakuin University, Tagajo} 
  \author{I.~Adachi}\affiliation{High Energy Accelerator Research Organization (KEK), Tsukuba} 
  \author{H.~Aihara}\affiliation{Department of Physics, University of Tokyo, Tokyo} 
  \author{D.~Anipko}\affiliation{Budker Institute of Nuclear Physics, Novosibirsk} 
 \author{K.~Arinstein}\affiliation{Budker Institute of Nuclear Physics, Novosibirsk} 
  \author{V.~Aulchenko}\affiliation{Budker Institute of Nuclear Physics, Novosibirsk} 
% ...
\collaboration{The Belle Collaboration}


\medskip
\date{January 4, 2007}


\maketitle

\tighten

{\renewcommand{\thefootnote}{\fnsymbol{footnote}}}
\setcounter{footnote}{0}


The nature of low mass (below 1~GeV/$c^2$) scalar mesons has been
a puzzle for decades with little progress made on its 
understanding~\cite{bib:scalar}.
Among the low mass scalar mesons, the existence of the $f_0\lr{980}$ and 
$a_0\lr{980}$ mesons is experimentally well established.
One of the key ingredients in understanding their nature is measurement of
the two-photon production cross sections and in particular
the two-photon widths extracted from them.
According to a relativistic quark model calculation, assuming the $f_0\lr{980}$
meson to be a non-strange $q \bar{q}$ state, its two-photon width 
should be in the range 1.3~keV to 1.8~keV~\cite{bib:relquark}.
However, a much smaller width is expected for an exotic state (0.2 - 0.6~keV
for a $K \bar{K}$ molecule state)~\cite{bib:barnes}, 
or for an $s \bar{s}$ state (0.3 - 0.5~keV)~\cite{bib:oller}.

A $B$ factory is one of the best laboratories for a detailed investigation
of low mass scalar mesons through two-photon production, where overwhelming
statistics can be obtained.
Two-photon production of mesons has advantages over
meson production in hadronic processes;
the production rate can be reliably calculated from QED 
with $\Gamma_{\gamma\gamma}$ as the only unknown parameter.
In addition, a meson can be produced alone without additional hadronic 
debris,
and the quantum numbers of the final state are restricted to states of 
charge conjugation $C=+1$ with $J=1$ forbidden
(Landau-Yang's theorem~\cite{bib:Yang}). 

In the past, using $209~{\rm pb^{-1}}$
of $\gamma\gamma\rightarrow\pi^+\pi^-$ data,
Mark II observed a shoulder in the $1~\GeV/c^2$ mass region,
which was tentatively identified as the $f_0\lr{980}$ 
resonance~\cite{bib:mark2}.
The reaction $\gamma\gamma\rightarrow\pi^0\pi^0$ was analyzed using
$97~{\rm pb}^{-1}$ of data taken with the Crystal Ball 
detector~\cite{bib:crysball}.
They found a hint of $f_0\lr{980}$ formation
with a significance of 2.2 standard deviations.
Measurements of $\gamma\gamma\rightarrow\pi^0\pi^0$ were also
performed with the JADE detector using $149~{\rm pb}^{-1}$
data~\cite{bib:JADE}.
They observed a small shoulder at around $1~\GeV/c^2$,
which was interpreted as the production of the $f_0\lr{980}$.
CELLO studied the reaction $\gamma\gamma\rightarrow\pi^+\pi^-$
using a data sample of 86~pb$^{-1}$ and concluded that an $f_0(980)$
signal at the level reported in 
Refs.~\cite{bib:mark2, bib:crysball, bib:JADE} cannot be excluded 
with their errors~\cite{bib:CELLO}.

Using data from Mark II, Crystal Ball, and CELLO,
Boglione and Pennington (BP) performed an amplitude analysis
of $\gamma\gamma\rightarrow\pi^+\pi^-$ and $\gamma\gamma\rightarrow\pi^0\pi^0$
cross sections~\cite{bib:Amplitude}.
They found two distinct classes of solutions where
one solution has a peak (``peak solution'') and the other has 
a wiggle (``dip solution'') in the $f_0\lr{980}$ mass region.
The two solutions give quite different results
for the two-photon width of the $f_0\lr{980}$ and
the size of the $S$-wave component.
Thus, it is important to distinguish them experimentally.

In this paper, we report on a high statistics study of the $f_0(980)$ meson
in the $\gamma\gamma\to\pi^+\pi^-$ reaction based on data taken with the 
Belle detector at the KEKB asymmetric-energy $e^+e^-$ 
collider~\cite{bib:kekb}.
The data sample corresponds to a total integrated luminosity of 85.9~fb$^{-1}$,
accumulated on the $\Upsilon(4S)$ resonance $(\sqrt{s} = 10.58~{\rm GeV})$
and 60~MeV below the resonance (8.6~fb$^{-1}$ of the total).
Since the difference in the cross sections  between the two energies is 
only about 0.3\%, we combine both samples.
We observe the two-photon process $e^+e^-\rightarrow e^+e^- \pi^+\pi^-$
in the ``zero-tag'' mode,
where neither the final-state electron nor positron is detected,
and the $\pi^+\pi^-$ system has small transverse momentum. 

A comprehensive description of the Belle detector is
given in Ref.~\cite{bib:belle}.
Charged track coordinates near the collision point are measured by a
silicon vertex detector (SVD) that surrounds a 2~cm radius beryllium
beam pipe.  
Track trajectory coordinates are reconstructed in a central drift chamber 
(CDC), and momentum measurements are made together with the SVD.
An array of 1188 silica-aerogel Cherenkov counters (ACC) provides separation
between kaons and pions for momenta above 1.2~GeV/$c$.  
The time-of-flight counter (TOF) system consists of a barrel-like arrangement
of 128 plastic scintillation counters and is effective for $K/\pi$ 
separation for tracks with momenta below 1.2~GeV/$c$. 
Low energy kaons and protons are also identified by specific
ionization ($dE/dx$) measurements in the CDC.
Photon detection and energy measurements of photons and electrons
are provided by an electromagnetic calorimeter (ECL) consisting of an 
array of 8736 CsI(Tl) crystals all pointing toward the interaction point.   
These detector components are located in a uniform magnetic field of 1.5~T
provided by a superconducting solenoid coil.
An iron flux-return located outside the solenoid coil is instrumented
to detect $K^0_L$ mesons and to identify muons (KLM).

Signal candidates are primarily triggered by a two-track trigger 
that requires two CDC tracks with associated TOF hits and ECL clusters 
with an opening angle greater than 135 degrees.
Exclusive $e^+e^-\to e^+e^-\pi^+\pi^-$ events are selected 
by requiring two oppositely charged tracks coming from the interaction 
region; each track is required to satisfy $dr<0.1$~cm and $|dz|<2$~cm, 
where $dr$ ($dz$) is the $r$ ($z$) component of the closest approach to the 
nominal collision point.
The $z$ axis of the detector is defined to be opposite to the direction
of the positron beam
and $r$ is the transverse distance from the $z$ axis.
The difference of the $dz$'s of the two tracks must satisfy the requirement
$|dz_+ - dz_-| \leq 1~{\rm cm}$.
The event must contain one and only one positively charged track
that satisfies $p_t > 0.3~{\rm GeV}/c$ and $-0.47 < \cos\theta < 0.82$,
where $p_t$ and $\theta$ are the transverse component of momentum and 
the angle with respect to the $z$-axis.
The scalar sum of track momenta in each event is required to be smaller than
$6~{\rm GeV}/c$,
and the sum of the ECL energies of the event must be less than
$6~{\rm GeV}$.
Events should not include an extra track with
$p_t > 0.1~{\rm GeV}/c$.
The cosine of the opening angle of the tracks must be greater than $-0.997$
to reject cosmic-ray events.
The sum of the transverse momentum vectors of the two tracks
$\lr{\sum \vec{p}_t^*}$ should satisfy
$\mid \sum \vec{p}_t^* \mid < 0.1~{\rm GeV}/c$;
this requirement separates exclusive two-track events from quasi-real 
two-photon collisions.

Electrons and positrons are distinguished from hadrons using 
the ratio $E/p$, where $E$ is the energy measured in 
the ECL, and $p$ is the momentum from the CDC.
Kaon (proton) candidates are identified using normalized kaon 
(proton) and pion 
likelihood functions obtained from the particle identification system 
($L_K$ ($L_p$) and $L_{\pi}$, respectively) 
with the criterion $L_K/(L_K+L_{\pi})>0.25$ ($L_p/(L_p+L_{\pi})>0.5$), 
which gives a typical identification efficiency of 90\% with a pion 
misidentification probability of 3\%.
All charged tracks that are not identified as electrons, kaons or protons are 
treated as pions.
We require both tracks to be pions.

In this measurement, the KLM detector cannot be used for muon identification,
since it is insensitive in the region of interest where the 
transverse momenta of tracks are below $0.8~\GeV/c$.
Therefore, we have developed a method
for statistically separating $\pi^+\pi^-$ and $\mu^+\mu^-$ events using 
ECL information; muons deposit energy corresponding to the 
ionization loss for minimum ionizing particles,
while pions give wider energy distributions since they interact 
hadronically in the ECL, which corresponds to approximately one interaction
length of material.
Probability density functions (PDFs) for the distributions of energy deposits
from $\mu^+\mu^-$ ($\pi^+ \pi^-$) pairs $P_{\mu^+\mu^-}\suplr{i} \lr{E_+,E_-}$
($P_{\pi^+\pi^-}\suplr{i} \lr{E_+,E_-}$)
are obtained with GEANT-3~\cite{bib:geant} Monte Carlo (MC) simulation.
Here $i$ represents the $i$-th bin of $(W, \abs{\cos\theta^*})$ 
in 20~MeV$/c^2$ and 0.1 steps,
where $W$ is the invariant mass of the $\pi^+ \pi^-$ (or $\mu^+ \mu^-$) 
pair in each event (the pion mass is assumed in the calculation), and
$\theta^*$ is the polar angle of the produced $\pi^\pm$ meson
(or $\mu^\pm$ lepton) in the center-of-mass system of two initial photons. 
Note that the effect of muons from pion decays is taken into account by 
the pion PDFs using this method.
We obtain $r\suplr{i}$, the fraction of $\mu^+ \mu^-$ in the $i$-th bin 
through the equation:
\begin{eqnarray}
N_{\rm data}^{\suplr{i}}\lr{E_+,E_-}
&=& N_{\rm tot}\suplr{i}  \left( 
 r\suplr{i} P_{\mu^+ \mu^-}\suplr{i} \lr{E_+,E_-}
\right. \nonumber \\ 
&& \left.  +
(1-r\suplr{i})P_{\pi^+ \pi^-}\suplr{i} \lr{E_+,E_-} \right) \; ,
\nonumber
\end{eqnarray}
where $N_{\rm data}^{\suplr{i}}\lr{E_+,E_-}$ is the distribution of data
and $N_{\rm tot}\suplr{i}$ is the total number of events in that bin.
The values of ratios $r\suplr{i}$ obtained must be corrected since the
MC cannot simulate hadronic interactions accurately enough.
By introducing mis-ID probabilities, 
$P_{\pi \pi \rightarrow \mu \mu}$ and $P_{\mu \mu \rightarrow \pi \pi}$, 
the $r$ value for each bin (the bin number $i$ is omitted) 
can be written as:
\begin{equation}
r = \frac{N_{\mu \mu} + N_{\pi \pi} P_{\pi \pi \rightarrow \mu \mu} 
- N_{\mu \mu} P_{\mu \mu \rightarrow \pi \pi}}{N_{\mu \mu} + N_{\pi \pi}}
\; ,
\nonumber
\end{equation}
where $N_{\mu \mu}$ ($N_{\pi \pi}$) is the number of true $\mu^+ \mu^-$ 
($\pi^+ \pi^-$) pair events in that bin.
We assume that $P_{\pi \pi \rightarrow \mu \mu}$ and 
$P_{\mu \mu \rightarrow \pi \pi}$ are independent of $W$.
Applying the $\mu / \pi$ separation method described above to a sample of data 
events positively identified as muons by the KLM, we find that 
$P_{\mu \mu \rightarrow \pi \pi}$ is statistically consistent with zero.
The values of $P_{\pi \pi \rightarrow \mu \mu}$ in each $|\cos \theta^*|$ bin 
are determined such that the ratio of the data and MC for
$\mu^+ \mu^-$ pairs, which is one ideally, gives a straight line in the $W$ 
spectrum. 
The values of $P_{\pi \pi \rightarrow \mu \mu}$ vary between 0.08 to
0.13 in $|\cos \theta^*|$ bins.
Because they are determined for each bin of $|\cos \theta^*|$,
the bin-by-bin variation of systematic errors is rather large in the angular
distribution.

The total cross section for $\gamma\gamma\to\pi^+\pi^-$ with
$|\cos \theta^*| < 0.6$ is evaluated using the following equation:
\begin{equation}
 \sigma_{\gamma\gamma\to\pi^+\pi^-}
  = \frac{\Diff N_{e^+e^-\to e^+e^-\pi^+\pi^-}}
         {\epsilon_{\rm trg}\cdot \epsilon_{\rm det} \cdot 
\Diff W\cdot \frac{d{\cal L}}{dW}\cdot \int L dt}.
\nonumber
\end{equation}
Here $\Diff N_{e^+e^-\to e^+e^-\pi^+\pi^-}$ is the number of events
in a $W$ bin, $\frac{d{\cal L}}{dW}$ is the two-photon luminosity 
function~\cite{bib:lum_func}
and $\int L dt = 85.9~{\rm fb}^{-1}$ is the integrated luminosity.
The size of the $W$ bin is chosen to be $5~\MeV/c^2$;
a typical mass resolution for a $\pi^+\pi^-$ system is 2~MeV$/c^2$
according to a MC study.
The detection (trigger) efficiencies, 
$\epsilon_{\rm det}$ ($\epsilon_{\rm trg}$)
are estimated with a MC simulation.
Events of the process $\gamma\gamma\to\pi^+\pi^-$ 
are generated using TREPS~\cite{bib:treps}.
The detection efficiency is extracted from MC simulation
and the trigger efficiency is estimated with the trigger simulator.
Since the trigger simulator does not simulate triggers well, particularly
in the low energy region, the efficiency values have to be corrected. 
We calculate the correction factors by comparing  $e^+e^-\to e^+e^-e^+e^-$ 
events in data and MC that are triggered by the two-track trigger.
The resulting factors steeply rise from 0.5 at $W=0.8~\GeV/c^2$ to 0.8 
at $W=1~\GeV/c^2$ and then increase gradually for higher $W$.
The muon-background subtraction and all the correction factors are applied 
using smooth functions obtained by parameterizing the results of bin-by-bin 
analyses.

\begin{figure}
 \centering
 \epsfig{file=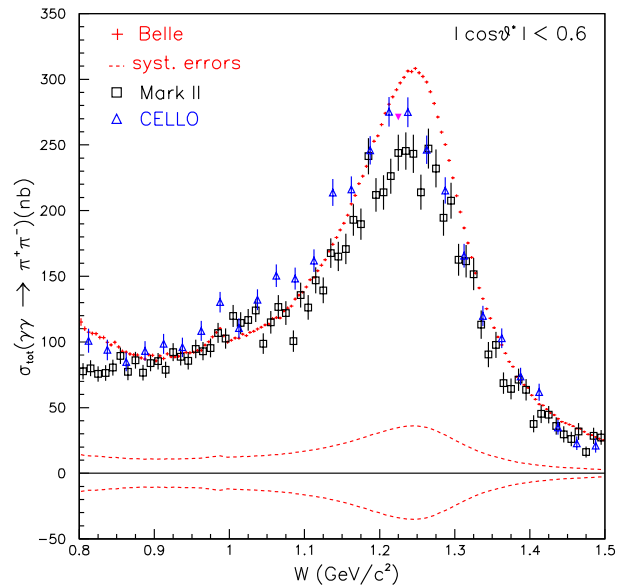,width=80mm}
 \centering
 \caption{The total cross section of $e^+e^-\to e^+e^-\pi^+\pi^-$ between
0.8 and 1.5~GeV/$c^2$ for $|\cos \theta^*|<0.6$. 
The Belle data are represented by crosses with statistical error bars, 
the CELLO data are the triangles and the Mark II data are squares. 
Dashed lines are upper and lower systematic
uncertainties for the Belle data.}
\label{fig:sigma_pipi_old}
\end{figure}

\begin{figure}
 \centering
 \epsfig{file=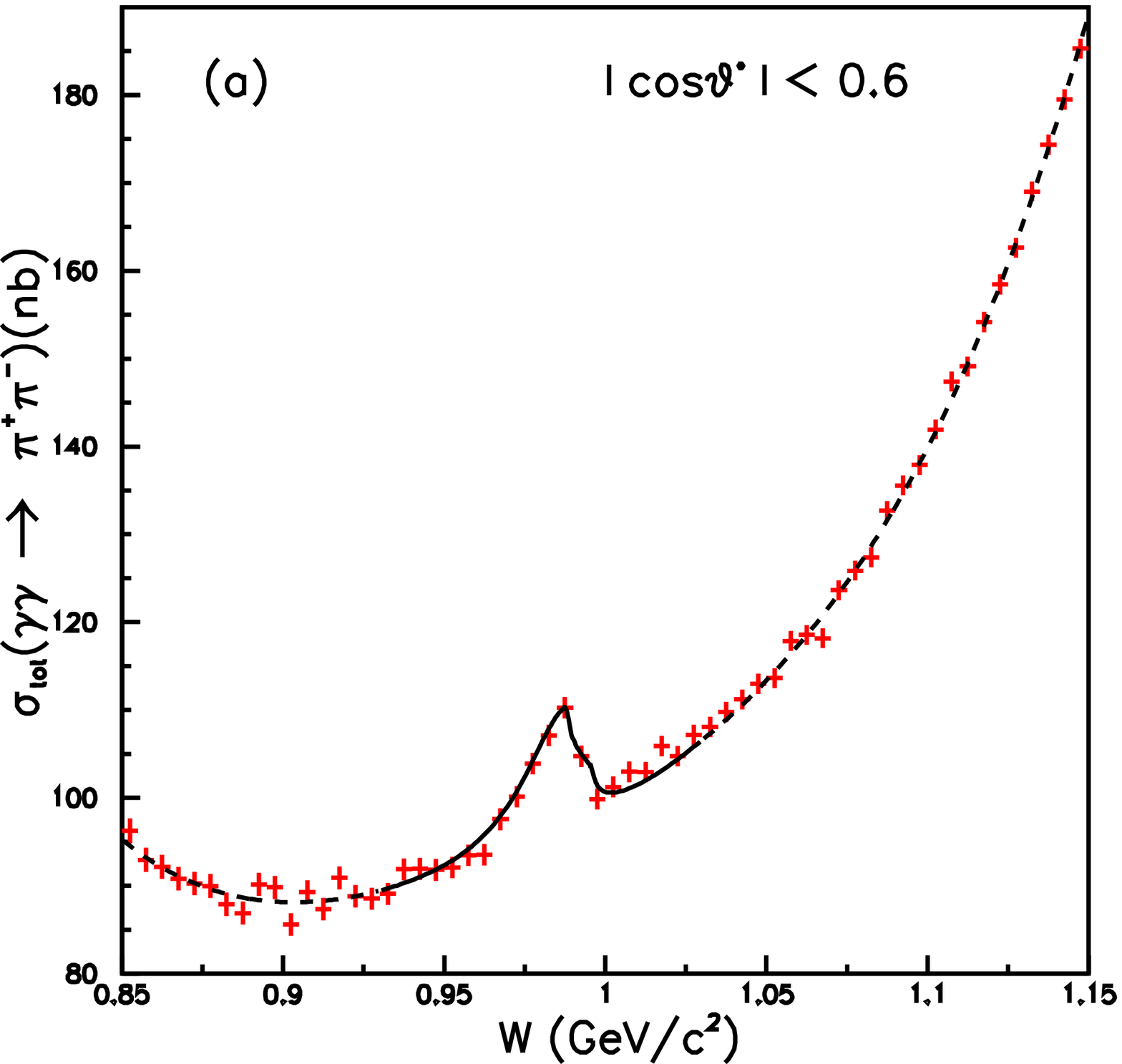,width=90mm}
 \epsfig{file=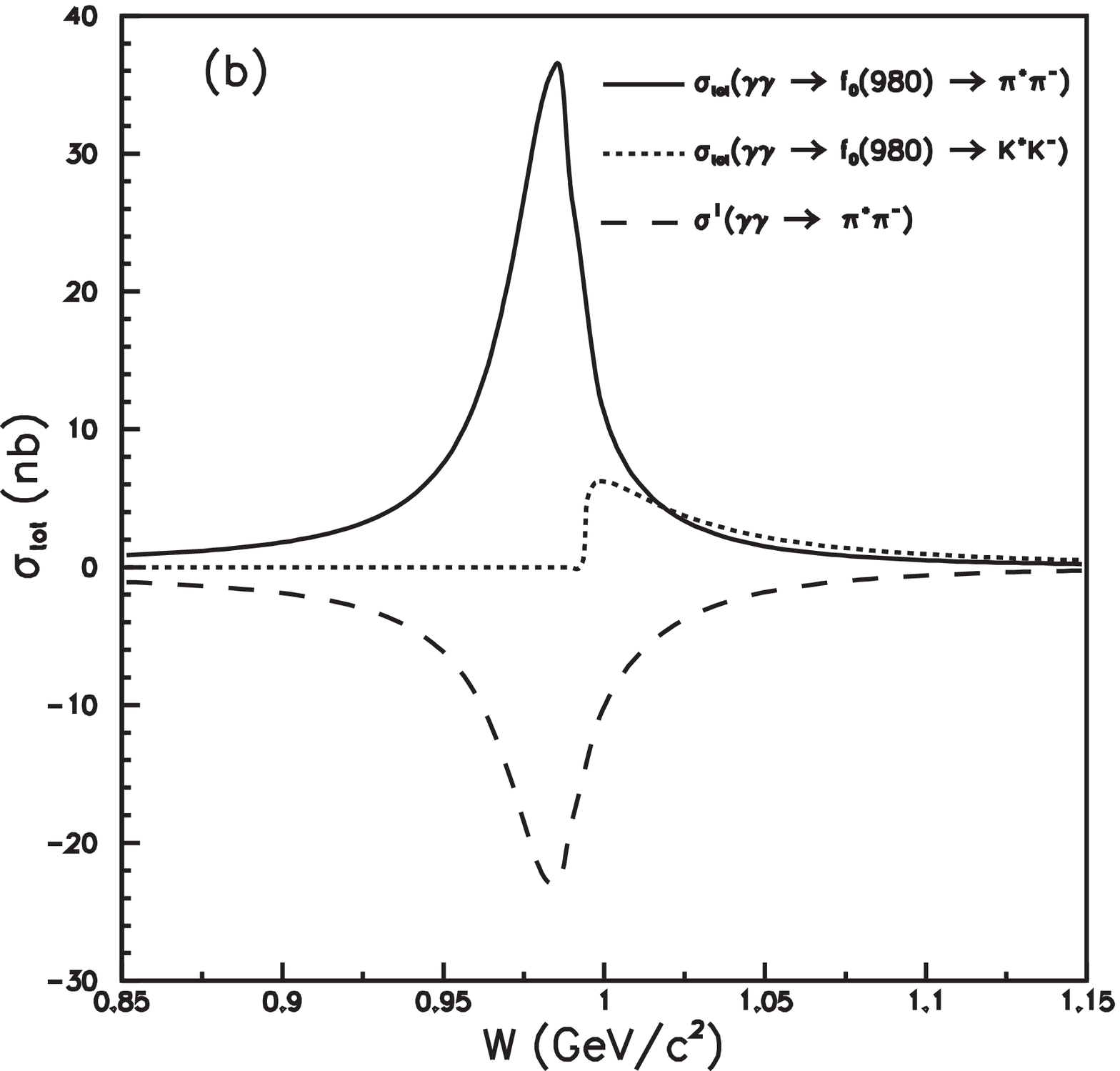,width=90mm}
 \centering
 \caption{Fitted curve: (a) shows the total cross section and
 (b) shows contributions of the resonance 
($\sigma(\gamma \gamma \rightarrow f_0(980) \rightarrow \pi^+ \pi^-))$
(solid line) and the interference (dashed).
The cross section of 
 $\sigma(\gamma \gamma \rightarrow f_0(980) \rightarrow K^+ K^-)$
is also shown (dotted).}
 \label{fig:f0_fit}
\end{figure}

The total cross section obtained is shown in Fig.~\ref{fig:sigma_pipi_old}
together with the results of some past experiments; an expanded
view of the $f_0(980)$ region is shown in Fig.~\ref{fig:f0_fit}(a).
A clear peak corresponding to the $f_0(980)$ meson is visible, and thus 
the peak solution of the BP analysis is selected. 
Systematic errors for the total cross section are summarized in 
Table~\ref{tab:syserr_sigma}.
They are dominated by the uncertainty of the $\mu/\pi$ separation and that
of the trigger efficiency.
Systematic errors arising from the $\mu/\pi$ separation are estimated by 
changing the value $P_{\pi \pi \rightarrow \mu \mu}$ in the allowable range 
in each angular bin. 
Since  $\mu^+ \mu^-$ events are well identified for $W > 1.6~\GeV/c^2$, the 
allowable range is determined in this region.
These well identified $\mu^+ \mu^-$ events are also used in estimating 
systematic errors of the trigger efficiency.
Comparing data and MC for $\mu^+ \mu^-$ events in the region 
$W > 1.6~\GeV/c^2$ and 
extrapolating linearly downward, the systematic errors are found to be 4\% 
at $W=1.5~\GeV/c^2$ and 10\% at $W=0.8~\GeV/c^2$.
The total systematic error is obtained by summing the systematic errors
in quadrature and is also shown in Fig.~\ref{fig:sigma_pipi_old}.
\begin{center}
 \begin{table}[h]
  \caption{Summary of systematic errors
           for the $\gamma\gamma\to\pi^+\pi^-$ cross section.
           A range is shown when the uncertainty has $W$ dependence.}
  \label{tab:syserr_sigma}
  \begin{tabular}{lc}
   \hline\hline 
   Parameter & Syst. error (\%) \\
   \hline
   Tracking efficiency   & 2.4     \\
   Trigger efficiency    & 4 -- 10  \\
   $K/\pi$-separation    & 0 -- 1       \\
   $\mu/\pi$-separation  & 5 -- 7   \\
   Luminosity function   & 5     \\
   Integrated luminosity & 1.4       \\
   \hline
   Total                 & 11.1 -- 12.3 \\
   \hline\hline 
  \end{tabular}
 \end{table}
\end{center}
\vspace{-\baselineskip}
Our results are in good agreement with past experiments
except for the $f_2\lr{1270}$ mass peak region, where they are
about 10 to 15\% larger, but still within the systematic errors.

A fit to the $\gamma\gamma\to\pi^+\pi^-$ total cross section
is performed to obtain the parameters of the $f_0\lr{980}$ meson.
We have to take into account the effect of the $K\bar{K}$ channel that opens
within the $f_0\lr{980}$ mass region.
The fitting function for the scalar resonance $f_0(980)$ is parameterized
as follows:
\begin{equation}
 \sigma = \abs{ \frac{\sqrt{4.8 \pi \beta_{\pi}}}{W}\F^{f_0}e^{i \varphi} +
 \sqrt{\sigma^{\rm BG}_0}}^2
  + \sigma^{\rm BG} - \sigma^{\rm BG}_0 ,
 \label{eqn:sigma}\\
\end{equation}
where the factor 4.8 includes the fiducial angular acceptance
$|\cos \theta^*|<0.6$,
$\beta_X = \sqrt{1-\frac{4 {M_X}^2}{W^2}}$ is the velocity of the 
particles $X$ with  mass $M_X$ in the two-body final states,
$\F^{f_0}$ is the amplitude of the $f_0(980)$
meson, which interferes with the helicity-0-background amplitude 
$\sqrt{\sigma^{\rm BG}_0}$ with relative phase $\varphi$, and 
$\sigma^{\rm BG}$ is the total background cross section.
The amplitude $\F^{f_0}$ can be written as 
\begin{equation}
\F^{f_0} = \frac{g_{f_0 \gamma\gamma}g_{f_0 \pi\pi}}{16\pi}
\cdot\frac{1}{D_{f_0}} ,
\label{eqn:ff0}
\end{equation}
where $g_{f_0XX}$ is related to the partial width 
of the $f_0(980)$ meson via
$\Gamma_{XX} (f_0) = \frac{\beta_X g_{f_0 XX}^2}{16 \pi M_{f_0}}$.
The factor $D_{f_0}$ is given as follows~\cite{bib:denom}:
\begin{eqnarray}
 D_{f_0}(W) &=& M_{f_0}^2 - W^2
              + \Re{\Pi_{\pi}^{f_0}}\lr{M_{f_0}}-\Pi_{\pi}^{f_0}\lr{W}
\nonumber \\
	    &&  + \Re{\Pi_K^{f_0}}\lr{M_{f_0}} - \Pi_K^{f_0}\lr{W} ,
\nonumber
\end{eqnarray}
where for $X = \pi$ or $K$,
\begin{equation}
 \Pi_X^{f_0}(W) =  \frac{\beta_X {g^2_{f_0XX}}}{16\pi}
       \left[i + \frac{1}{\pi}
			          \ln\frac{1-\beta_X}
				          {1+\beta_X}\right] .
\end{equation}
The factor $\beta_K$ is real in the region $W \geq 2M_K$
and becomes imaginary for $W < 2M_K$.
The mass difference between $K^{\pm}$ and $K^0$ $(\overline{K^0})$ is 
included by using $\beta_K = \frac{1}{2} (\beta_{K^{\pm}} + \beta_{K^0})$.

In the fit, we assume $\sigma^{\rm BG}_0$ to be constant 
and the relative phase to be a slowly varying function of $W$; this
is motivated by the nearly energy-independent behavior of the scalar
Born amplitude~\cite{bib:morgan}.
We fix $g_{f_0KK}^2 / g_{f_0\pi\pi}^2 = 4.21\pm 
0.25\lr{\rm stat}\pm 0.21\lr{\rm syst}$ taking the latest
value from the BES measurement~\cite{bib:bes}. 
The background function $\sigma^{\rm BG}$ is evaluated by
fitting the cross section with a 4-th order polynomial in $W$ outside of the 
$f_0\lr{980}$ region 
$0.85~\GeV/c^2 < W < 0.93~\GeV/c^2$ and $1.03~\GeV/c^2 < W < 1.15~\GeV/c^2$.
The value of $\chi^2 /ndf$ for the fit is 0.88 for 46 degrees of 
freedom ($ndf$).
A fit to the $f_0\lr{980}$ resonance is then performed
with Eq.~(\ref{eqn:sigma}) in the region
$0.93~\GeV/c^2 < W < 1.03~\GeV/c^2$, where the parameters of $\sigma^{\rm BG}$
are fixed; the free parameters
are $M_{f_0}$, $g_{f_0 \pi \pi}$, $\Gamma_{\gamma \gamma}$ (evaluated at
the $f_0(980)$ mass), $\sigma^{\rm BG}_0$ and $\varphi$.

The result of the fit is shown in Figs.~\ref{fig:f0_fit}(a) and 
\ref{fig:f0_fit}(b).
In Fig.~\ref{fig:f0_fit}(b), one can see a significant interference effect,
which is visible as a deviation from a Breit-Wigner-like shape
in Fig.~\ref{fig:f0_fit}(a). 
In the same figure, the cross section
 $\sigma(\gamma \gamma \rightarrow f_0(980) \rightarrow K^+ K^-)$
is also plotted, which is obtained by evaluating the first term  in 
Eq.~(\ref{eqn:sigma}), substituting $\beta_K$ instead of
$\beta_{\pi}$ and in Eq.~(\ref{eqn:ff0}) $g_{f_0 K K}$ instead of
$g_{f_0 \pi \pi}$.
Note that the cross section is zero below the threshold even though
the amplitude is non zero.
The value of $\chi^2 /ndf$ of the fit is 1.04 for 15 $ndf$.
The helicity-0-background component that interferes with the $f_0(980)$ meson 
($\sigma^{\rm BG}_0$) is found to be $3.7^{+1.2}_{-0.5}$~nb.
The value of $\varphi$ is approximately $\pi/2$, which is consistent with the 
general phase shift study~\cite{bib:Amplitude}.

The parameters of the $f_0\lr{980}$ meson are found to be
\begin{eqnarray}
 M_{f_0} &=& 985.6 ~^{+1.2}_{-1.5}\lr{\rm stat}
                   ~^{+1.1}_{-1.6}\lr{\rm syst}~\MeV/c^2
\nonumber \\
 \Gamma_{\pi^+\pi^-}\lr{f_0} &=& 34.2~^{+13.9}_{-11.8}\lr{\rm stat}
                             ~^{+8.8}_{-2.5}\lr{\rm syst}~\MeV \nonumber \\
 \Gamma_{\gamma\gamma}\lr{f_0} &=& 205
                                   ~^{+95}_{-83}\lr{\rm stat}
                                   ~^{+147}_{-117}\lr{\rm syst}~\eV.\nonumber 
\end{eqnarray}
The two-photon width given by the PDG~\cite{bib:PDG} is
$ \Gamma_{\gamma\gamma}\lr{f_0} = 310~_{-110}^{+80}\lr{\rm stat}~\eV$, and
the value found by BP is $280^{+90}_{-130}$~eV.
Our results are consistent with them within errors as well as
with the prediction of the four-quark model of 270~eV~\cite{bib:achasov}.

The dominant systematic errors come from fitting.
The value of $ \Gamma_{\gamma\gamma}\lr{f_0} $ is quite sensitive
to changes in parameters of the background cross section (fitted 
outside of the $f_0(980)$ resonance).
Systematic errors are evaluated by changing each background parameter
by $\pm 1 \sigma$, taking their correlations into account;
the error is strongly correlated with that of $g_{f_0 \pi\pi}$
(i.e. $\Gamma_{\pi^+ \pi^-}\lr{f_0}$).
The error in the normalization of the total cross section has little effect
on the value of the $f_0(980)$ mass, however it is a significant contribution
 to the error in 
$ \Gamma_{\gamma\gamma}\lr{f_0}$ and $ \Gamma_{\pi^+ \pi^-}\lr{f_0}$.
The errors in $g_{f_0KK}^2 / g_{f_0\pi\pi}^2$ are also taken into account
in the systematic errors.
Individual systematic errors are summed in quadrature to obtain the total
uncertainty.

In summary, we have made a high statistics measurement of
the $\gamma \gamma \rightarrow \pi^+ \pi^-$ cross section in the
$\pi^+\pi^-$ invariant mass region $0.80~\GeV/c^2 \leq W \leq 
1.5~\GeV/c^2$
in fine bins of $W$ (5~MeV) and $\cos\theta^*$ (0.05)
with the Belle detector at the KEKB accelerator.
We have observed a significant signal corresponding to the $f_0\lr{980}$ 
resonance.
Our data clearly select the peak solution of the Boglione-Pennington
amplitude analysis~\cite{bib:Amplitude}.
The total cross section is fitted to obtain the parameters of the
$f_0(980)$ meson.
Its two-photon width is found to be $205 ~^{+95}_{-83}\lr{\rm stat}
                                   ~_{-117}^{+147}\lr{\rm syst}$ eV,
consistent with past experiments. 

We are indebted to T.~Barnes who provided us with a more complete list of 
theoretical references to calculations of the two-photon widths of scalar 
mesons, and to M.~Pennington for various enlightening discussions and useful 
suggestions.
We thank the KEKB group for excellent operation of the
accelerator, the KEK cryogenics group for efficient solenoid
operations, and the KEK computer group and
the NII for valuable computing and Super-SINET network
support.  We acknowledge support from MEXT and JSPS (Japan);
ARC and DEST (Australia); NSFC and KIP of CAS (China); 
DST (India); MOEHRD, KOSEF and KRF (Korea); 
KBN (Poland); MIST (Russia); ARRS (Slovenia); SNSF (Switzerland); 
NSC and MOE (Taiwan); and DOE (USA).

\end{document}